\documentclass[12pt,a4papper]{article}
\usepackage{graphicx}
\usepackage{amssymb}\usepackage{amsmath}

\usepackage{color}

\begin{document}

\title{The dynamical nature of time}
\author{Antonio F. Ra\~nada\thanks{Facultad de F\'{i}sica, Universidad Complutense, 28040 Madrid, Spain. Email:
afr@fis.ucm.es}
 \and A. Tiemblo\thanks{Instituto de Matem\'aticas y F\'{i}sica Fundamental, Consejo Superior de
 Investigaciones Cient\'{i}ficas, Serrano 113b, 28006 Madrid, Spain.
 E-mail: Tiemblo@imaff.cfmac.csic.es}}
\date{19 July 2011}

\maketitle

\begin{abstract} It is usually assumed that the ``$t$'' parameter in the
equations of dynamics can be identified with the indication of the
pointer of a clock. Things are not so simple, however. In fact,
since the equations of motion can be written in terms of $t$ but
also of $t'=f(t)$, $f$ being any well behaved function, any one of
those infinite parametric times $t'$ is as good as the Newtonian
one to study classical dynamics in Hamiltonian form. Here we show
that, as a consequence of parametric invariance, the relation
between the mathematical parametric time $t$ in the equations of
dynamics and the physical dynamical time $\sigma$ that is measured
with a particular clock (which is a dynamical system) is more
complex and subtle than usually assumed. These two kinds of time,
therefore, must be carefully distinguished. Furthermore, we show
that not all the dynamical clock-times are necessarily equivalent
and that the observational fingerprint of this non-equivalence
has, curiously, the same form as that of the Pioneer anomaly, a
still unexplained phenomenon.

\end{abstract}

\section{Introduction}

The main problem of dynamics is probably to understand in depth
the role and meaning of the term ``time''. Two kinds of time are
used in physics. On one side, the parametric time $t$, just an
auxiliary mathematical element which, strictly speaking, is not
observable since any other time of the form $t'=f(t)$, $f$ being
any well behaved function, serves equally to describe  the motion
of a dynamical system. On the other the time measured with
particular clocks, say $\sigma$, which are dynamical systems
obeying the laws of physics.  The latter is a dynamical variable,
for instance the angle of a pointer, and deserves therefore to be
qualified as dynamical.  The consequences of the existence of
these two kinds of time, parametric and dynamical, have not been
perhaps explored enough.

Here we show that the dynamical time $\sigma(t)$ measured by a
clock $\sigma$, can be obtained as the solution of the equation of
motion that characterizes the clock, of the form ${\rm d} \sigma
/{\rm d} t=u(t)$, where $u(t)$  denotes here the ``march" of the
clock $\sigma$ with respect to the parametric time $t$. While
$\sigma(t)$ has a real dynamical character, $t$ is just a
mathematical parameter, which (i)  has a purely auxiliary role to
write the action and obtain the equations of motion, (ii) lacks
any physical or dynamical nature,  iii) it is a symbol that
describes the evolutive character of the reality and (iv) is not
observable. The differences between parametric and dynamical times
could have significant consequences, since two dynamical
clock-times, say $\sigma_1$ and $\sigma_2$, are not necessarily
equivalent, so that there could be different times accelerating
with respect to one another. The consequences of these arguments
could be important; we just mention here two cases in which they
could shed some light. First is the meaning of the cosmic time.
Second, the fourth Heisenberg relation which requires that the
time be a dynamical variable.

In order to write the equations of motion of a system in terms of
really observable and dynamical quantities, what is done is to compare two
motions, one of the system and the other of a standard clock. This
requires the use of two principles. The first is the
parametric invariance under transformations $t\rightarrow t'=f(t)$, an important property of gravitation theories, the other is a principle of coherence, {\it
i.e., } that the equations of motion of both the system and the
clock be described by the same physical theory.\\

\section{Parametric invariance in classical dynamics}
Though the parametric time is a fundamental concept in classical
dynamics, as said before, it has a non-dynamical character. As a
consequence, there is no
 canonical momentum conjugate to $t$. Common wisdom assumes that this non-dynamical $t$ is measured with a
 clock, but this assumption must be submitted to a rigorous analysis.
 Note that it is always possible to synchronize two clocks at a certain initial time
 $t_0$, but what cannot be assured is that they will keep ticking
 at the same rate. This raises the question whether the
 equations of motion of dynamics depend or not on the march of the clocks,
 which implies the need to establish a parametric invariance principle.
 There exists a scheme in which all these problems can be solved by means of the introduction
 of the idea of a dynamical time \cite{Han76,RT08}. The $t$ variable appearing in this scheme is
 just a non-observable  auxiliary parameter. In fact, the theory
 so constructed is parametric invariant, as happens also in general relativity.

In order to do that we replace the standard action $S=\int
[p\,\dot{q}-H(p,q)]{\rm d} t$, by the alternative expression
\begin{equation} S=\int\{\Pi
(t)\dot{\sigma}_0(t)+p\,(t)\dot{q}(t)-u(t)[ \Pi(t
)+H(p\,(t),q(t))]\}{\rm d} t \,,\label{20}\end{equation}
 (overdot means derivation with respect to the auxiliary parameter $t$), where $\sigma_0(t)$,
$\Pi(t)$ are conjugate variables that describe the behavior of the
clock, and $\Pi _u$,  the momentum conjugate to $u(t)$, weakly
vanishes.

The corresponding Hamiltonian is  $\hat{H}=u[\Pi +H(p,q)]+\lambda
\,\Pi _u$ where $\lambda$ is a Lagrange multiplier. The stability
of the weak condition $\Pi _u=0$ implies the following first class
constraint  \begin{equation} \Pi
+H(p,q)=0,\label{25}\end{equation}

 which induces the following reparametrization
transformations $\delta \sigma _0=\alpha (t)$,
 $\delta
q=\alpha (t) \dot{q}$ and $\delta p=\alpha (t) \dot{p}$ with
$\alpha(t)$ being an arbitrary function.

 The
transformations induced by $\Pi _u$ allow then to interpret $u(t)$
as an arbitrary function, so that the Hamiltonian becomes
\begin{equation} H^E =u[\Pi +H(p,q)]\,,\label{40}\end{equation}
Though this Hamiltonian reduces to a first class constraint, it
contains a very realistic dynamical evolution, given by the
Hamiltonian equations
 \begin{equation} \dot{q}=u{\partial H\over
\partial p},\quad  \dot{p}=-u{\partial H\over \partial q},\quad
u=\dot{\sigma}_0,\quad
\dot{\Pi}=-\dot{H}=0.\label{50}\end{equation} It follows that
\begin{equation} {{\rm d} q\over {\rm d} \sigma_0}={\partial H\over
\partial p},\quad {{\rm d} p\over {\rm d} \sigma_0}=-{\partial H\over
\partial q},\quad u={{\rm d} \sigma_0 \over {\rm d} t},\quad {{\rm d}
H\over {\rm d} \sigma_0}=0\,, \label{60}\end{equation} equations
that are full of dynamical significance. The first two are the
canonical equations of motion with the dynamical time $\sigma_0$
as the time variable, in such a way that the evolution becomes a
correlation between dynamical variables. The third one can be
interpreted as the equation of motion ({\it i.e., } the ``march'')
of $\sigma_0$ with respect to the parameter $t$. Notice that the
total Hamiltonian $\hat{H}=u[\Pi +H(p,q)]+\lambda \,\Pi _u$ is the
sum of two terms, describing, respectively, the physical system
and the clock. The equation of motion of the second term,
$H_{\rm clock}=u\Pi + \lambda \Pi _u$, is precisely that of a
clock $u={\rm d}\sigma_0 /{\rm d}t$.

Since this theory is invariant under reparametrization, we may
fix, for instance, the gauge by the condition $\sigma_0 =t$ ({\it
i.e., } $u=1$), so that we recover the ordinary canonical formalism
with $t$ being the Newtonian time. Notice that the choice of gauge
means in fact to choose a clock.

The observations are performed with real clocks, which are
dynamical systems, each one with a dynamical variable that is a
well behaved increasing function of $t$ and can therefore be
identified with a dynamical clock-time $\sigma(t)$, which can be
used to fix the reparametrization gauge. As long as the
observations make use of only the standard dynamical clock
$\sigma_0$, the scheme is nothing else than the Hamiltonian
equations. This may not occur, however, if a real clock
$\sigma(t)$ with a different march is involved. In the latter
case, the motion equations are (\ref{60}), but with $\sigma$ and
$\sigma _0$ instead of $\sigma_0$ and $t$, respectively,
\begin{equation} {{\rm d} q\over {\rm d} \sigma}={\partial H\over
\partial p}\,;\qquad {{\rm d} p\over {\rm d} \sigma}=-{\partial H\over
\partial q}\,;\qquad u={{\rm d} \sigma \over {\rm d} \sigma _0}\,, \label{70}\end{equation}
which describe the physics of a system in operationally realistic
terms. This means that they do not refer to any unobservable
parametric time but to $\sigma$, which is the time really observed
by a real clock. The novelty is here the presence of the third
equation (\ref{70}), which is the dynamic equation of the second
clock with respect to the standard one. The important fact for our
purposes is that classical dynamics can be formulated as a
parametrically invariant theory.

\section{The relativistic particle}
 Before going into this section let us summarize the arguments of the
 previous one. Starting from a Hamiltonian theory with $n$ degrees
 of freedom, we introduced a new one (the dynamical time), in such a way that the
 motion equations become correlations between dynamical variables only (\ref{60}).
 Nevertheless the new theory has a first class constraint
 (reparametrizations) that allows us to fix arbitrarily the value of
 $\sigma_0$. The only way to do that is to choose a new dynamical
 system, {\em i.e., } a real clock such as the Earth's motion or any
 other, with a well behaved dynamical variable $\sigma (t)$ as
 appears in (\ref{70}). So in practice to fix the gauge of
 the symmetry under reparametrizations is to choose a clock. In
 other words, we measure a motion using another one as a standard.
It must be underscored that, to completely formulate the equations
of a dynamical system, the chosen clock must be specified.

 The kinematics of the  free particle in special
relativity follows the same scheme. The parametric invariant
action ${\cal S}=mc\int {\rm d} s$, corresponds to the Lagrangian
(overdot means derivative with respect to an arbitrary time)
\begin{equation} {\cal
L}=-mc\sqrt{\dot{x}_0^2-\dot{x}_1^2-\dot{x}_2^2-\dot{x}_3^2}
\label{100}\end{equation}

It is easy to see that there is a first class primary constraint
of the form \begin{equation}
p_0^2-p_1^2-p_2^2-p_3^2=m^2c^2,\label{110}\end{equation} that
expresses the evident parametric invariance of the action. Due to
the existence of this primary constraint not all the time
derivatives of the coordinates can be obtained in terms of the
momenta. Choosing now $\dot{x}_0$ as an arbitrary function of $t$,
the Hamiltonian becomes \begin{equation}
H^E=\dot{x}_0\left(p_0+\sqrt{p_1^2+p_2^2+p_3^2+m^2c^2}\right),
\label{110}\end{equation} which reproduces (\ref{40}) with $p_0$
playing the role of $\Pi$ and the square root being $H(p,q)$. It
is clear thus that $x_0(t)$ must be interpreted as a dynamical
time.

Let us take now the motion of a particle in a general metric
tensor $g_{\alpha\beta}$, so that ${\rm d}
s=\sqrt{g_{\alpha\beta}{\rm d} x^\alpha {\rm d} x^\beta}$. The
Lagrangian is \begin{equation} {\cal L}=
-mc\sqrt{g_{\alpha\beta}\dot{x}^\alpha \dot
{x}^\beta}.\label{120}\end{equation} Note that, since the motion
is geodesic, the components of the metric tensor are not dynamical
variables but prescribed functions of the coordinates. Following
the same procedure as in the previous case, the primary constraint
is now $g^{\alpha\beta}p_\alpha p_\beta=m^2c^2$. Thus the
Hamiltonian becomes \begin{equation}
H=\dot{x}^0[p_0-(N\sqrt{p_ip^i+m^2c^2}+p_iN^i)],\label{140}\end{equation}
where $i=1,2,3$, $N$ is the lapse, $N_i$ the shift and the Latin
indices are raised and lowered with the three-dimensional metric.
 As we see, the situation is
the same as in previous cases, $x^0$ playing the same role as the
dynamical time.

 It must be underlined that all the previous
Hamiltonians are, in fact, first class constraints. They generate,
however, well defined dynamical evolutions (see
(\ref{60})--(\ref{70})). Notice that they contain two terms that
describe i) the dynamical system which is studied and ii) a
particular clock.

The case of the particle in a gravitational field
$g_{\alpha\beta}$ illustrates the difference between the spatial
coordinates $x^i$ and the temporal one $x^0$ since the former can
be chosen arbitrarily while the latter needs an additional
dynamical system (a real clock) in order for it to be fixed so
that probably a (3+1)-spacetime is closer to the reality than a
4-spacetime.

\section{The Einstein--Hilbert action}

General relativity was constructed to be a parametric invariant
theory from its very foundation, as happens with any other
diff-invariant theory. Its essential difference from the previous
examples is that, in the former cases, the dynamical variables are
the coordinates, defined in a non-dynamical metric. Conversely, in
the latter,  the dynamical variables are the components of the
metric tensor, while the coordinates are auxiliary objects with no
dynamical nature. Accordingly to our previous statements, we will
take from now on a (3+1)-spacetime. In the ADM scheme \cite{ADM59}
the Hamiltonian becomes \begin{equation} H^E= \int {\rm d} ^3x
[N{\cal H}(q_{ij},\,\pi^{ij})+N_i {\cal \chi}^i
(q_{ij},\,\pi^{ij})],\label{160}\end{equation} where $N$ and $N_i$
are the lapse and the shift, respectively, $q_{ij}$ the 3-metric
and $\pi ^{ij}$ its canonically  conjugate momentum. The absence
of time derivatives of $N$ and $N_i$ determines  the presence of
primary first class constraints, which implies in turn that $N$
and $N_i$ are arbitrary functions. The secondary first class
constraints ${\cal H} =0$ and $\chi _i =0$ fix the subspace in
which the motion takes place. If one fixes  $N_i=0$, the
Hamiltonian becomes $H=\int {\rm d} ^3x\, N{\cal H}$.

From this expression one could reproduce the same process followed
before in the case of ordinary analytical dynamics. To interpret
the dynamics described by a Hamiltonian such as (\ref{160}) it
suffices, maintaining $N_i=0$, to consider the meaning of $N$,
defined as ${\rm d} \tau/{\rm d} t$ where ${\rm d} \tau$ is the
proper time distance between two shells of the foliation.  Note
that $N$ is an arbitrary dynamical variable which plays thus the
same role as $\dot{x}^0$ and $\dot{\sigma}$ in the previous cases:
all of them are derivatives with respect to the parametric time.
We find, therefore, that the dynamical time coincides with the
proper time. Nevertheless, as is suitable to general relativity,
the dynamical time is just a local time.

Fixing $N=1$ implies the use of proper time. In the case of the
electromagnetism equations in a gravitational field, a geometrical
contribution to the permittivity and permeability appear which
modify the values of $\varepsilon_0$ and $\mu_0$ and thus the speed
of light. This change is easily avoided by using the proper time
as the dynamical time \cite{Lan75}. In any case, the speed of light  is
still a fundamental constant if measured with atomic clocks since
the periods of atomic oscillations are obviously constant with
respect to it. Curiously, atomic clocks measure proper time,
notwithstanding the fact that they are quantum devices described
by quantum physics, while the proper time is a classical concept.

The choice of a physical clock is then a most relevant question.
The clock must comply with some obvious conditions. It must be a
dynamical system, the solution of its equation of motion $\sigma
(t)$ being a well behaved and monotonously increasing function of
the parametric time $t$, as for instance the number of cycles of
an harmonic oscillator or of the Earth rotation. Strictly speaking
the fixing of a gauge is a mathematical question, though physically
relevant since it is equivalent to the choice of a clock. It must
be underscored that the complete description of a dynamical system
needs to specify the clock which is used. This is a very important
problem, specially for cosmological models.

It must be underscored that the previous arguments imply that the
parametric invariance is the main characteristic of classical
dynamics. {\it I.e.}, this invariance states  that the equations of motion
are independent of the clock used to observe the
trajectory. Otherwise said, it is a way to restrict to the time
variable the principle of general covariance of relativistic
physics.

 Let us see what would happen if parametric invariance is not taken into account. For this purpose and in order to understand general relativity,
 simplified models have
been proposed to obtain valuable information in areas such as quantum
gravity or cosmology. The usual strategy is to kill some degrees
of freedom. There is a way, however, to achieve the same result
but going in the opposite direction, {\it i.e., } adding degrees of
freedom. This is the case of the Husain--Kucha$\check{\mbox{r}}$
model \cite{Hus90}, which lacks the Hamiltonian (scalar
constraint) in such a way that the number of degrees of freedom
per space point grows from 2 to 3. In such a theory, parametric
invariance would be absent. The price to be paid then is that the
four dimensional metrics that can be constructed seem to be
degenerate. Without discussing this point here, it is important to
state that the Husain--Kucha$\check{\mbox{r}}$ model is a
particular case of a more general theory (see \cite{Bar98} for
details) that includes nondegenerate metrics if a dynamical time
variable is present.

\section{A principle of coherence}
As was pointed out at the end of Section 1, when two clocks are
involved the question of their coherence must be considered. There
is no problem if the dynamics of both the system and the clock are
governed by the same physical theory. This is because any
discrepancy between two clocks must be solvable, from the
theoretical point of view, in the frame of the theory itself. For
instance, the effect of the tides on the Earth's rotation modifies
the value of the day, an effect that can be calculated by taking
into account the gravitation involved in the Earth--Moon system.

This requirement of coherence, which guarantees that the equation
of motion of the clock ({\it i.e.}, its march) is given by the
same theory as that of the dynamical system, cannot be maintained
when the clock and the system  obey two different theories. This
is the case when atomic clocks are used in classical general
relativity. Lacking a quantum gravity theory, the equation of
motion of the atomic clocks $\sigma_2(t)$ cannot be determined
{\it a priori} and, consequently, it is not possible to compare it
with the equation of motion $\sigma_1(t)$ of a classical clock.
The only way to do so relies necessarily on empirical methods.
Note that if it is found that the two marches are different, this
does not necessarily imply a violation of parametric invariance.

The previous considerations certainly clarify the role of the
clocks and the meaning of the word ``time''. The two main kinds of
clocks used in physics are the astronomical and the atomic ones,
which are dynamical systems based on classical and quantum
physics, respectively. The solar system taken as a clock gives the
ephemeris time while the vibrations of quantum systems measure the
atomic one. Current wisdom assumes implicitly that these two types
of clocks give the same time but, as explained before, this is not
necessarily so. Indeed there is no {\it a priori} reason to
postulate that two clocks beat at the same rate if they are based
on two different theories, such as gravitation and quantum physics
which are not only different but, even more, all
efforts to unify them have failed up to now.

\section{Looking for observational evidence}

The arguments of this paper show that the difference between
$t_{\rm astr}$ and $t_{\rm atom}$ is either nil or very small,
otherwise an unexpected new effect should have been detected by
now. Let us admit that it is non-nil. Because of the continuous
improvement of measurement devices during the last decades, an
observational test of the relative acceleration between these two
times might already be available, although we could be unaware of
this possibility. What's more, the effect could have been observed
by now but without being properly interpreted.  A provoking case
could be a spaceship receding from the Sun. Since its trajectory
is calculated with standard gravity theories that use astronomical
time but it is measured with devices based on quantum physics that
use atomic time, some anomaly could be observed. In fact the
theory gives the ship's trajectory as a certain function
parametrized by astronomical time ${\bf r}={\bf r}(t_{\rm astr})$
but the observations see the same three-dimensional trajectory,
although parameterized by atomic time and given by a different
function ${\bf r}'={\bf r}'(t_{_{\rm atom}})$. The two times are
related as ${\bf r}'(t_{\rm atom})={\bf r}(t_{\rm astr})$ (they
are examples of the aforementioned clock-times $\sigma_2(t)$ and
$\sigma_1(t))$. It is clear that they can be synchronized at a
certain initial time so that $t_{\rm astr,\,0}=t_{\rm
atom,\,0}=t_0$, but they will start to desynchronize progressively
afterwards as \begin{equation} {\rm d} t_{\rm
atom}=[1+a(t-t_0)]\,{\rm d} t_{\rm astr},\quad \mbox{ with }\qquad
a={{\rm d} ^2t_{\rm atom}\over {\rm d} t_{\rm astr}^2},
\label{180}\end{equation} where the small inverse time $a$ is the
relative acceleration of $t_{\rm atom}$ and $t_{\rm astr}$, and
$u={\rm d} t_{\rm atom} /{\rm d} t_{\rm astr}=1+a(t-t_0)$ the
march of $t_{\rm atom}$ with respect to $t_{\rm astr}$. Note that
it is not necessary, at first order, to specify which one of the
two times is $t$.

Defining the velocities of a mobile with respect to the two times
as $v_{\rm atom} ={\rm d} \ell/{\rm d} t_{\rm atom}$ and $v_{\rm
astr} ={\rm d} \ell/{\rm d} t_{\rm astr}$, it follows that
\begin{equation} v_{\rm atom} = {v_{\rm astr}\over u}, \qquad \qquad
{\Delta v\over v}= -a(t-t_0), \label{190}\end{equation} with
$\Delta v =v_{\rm atom} -v_{\rm astr}$.
  As
could have been expected, the observational fingerprint of the
relative acceleration of the two clock-times  would be a
discrepancy between the expected and observed speeds of a mobile.
This implies that the speed of light would depend on which
clock-time is used: it is a fundamental constant only if measured
with atomic clock-time. It must be so since the periods of the
atomic oscillations are obviously constant with respect to $t_{\rm
atom}$, in fact they are its basic units (see \cite{RT08,Ran04}
where the details are explained). Note that
(\ref{180})--(\ref{190}) imply that if $a<0$, then $v_{\rm atom}
>v_{\rm astr}$ while if $a>0$, then $v_{\rm atom} <v_{\rm astr}$ (assuming $t>t_0$). In the latter
case, the ship would seem to lag behind the position predicted by
gravitation theories.

In fact quite a similar lag has already been observed and has even
a name: the Pioneer anomaly. Surprisingly, it remains unexplained
more than thirty years after being discovered by Anderson {\it et
al.} in 1980 \cite{And98,And02,Tur10b} in spite of many efforts to
account for it. What is important here is that the observational
fingerprint of the anomaly has the same form as the second
equation (\ref{190}). What Anderson {\it et al.} found is that the
frequencies of the two-way signals to and from the Pioneer 10
spaceship included an unexpected Doppler residual which did not
correspond to any known motion of the ship.  They were able to
measure the value $a =(5.84\pm 0.88)\times 10^{-18}\mbox{
s}^{-1}$, although using the inverse time $a_{\rm t}=a/2$, and
suggested that $a_{\rm t}$ could be ``like a non-homogeneity of
time''\  or a ``clock acceleration'' \cite{And98}. But they did
not explain acceleration with respect to what, nor  did they
develop any theoretical analysis of this idea, assuming at first
that $2a_{\rm t}$ was just the measure of a real Doppler effect.
However it was soon understood that this interpretation is neither
compatible with the equivalence principle nor with the cartography
of the solar system. For several years it was thought that
systematics would be the most probable explanation of the anomaly
(see the conclusions of \cite{And02}) but no error was found in
spite of several different mathematical analyses of the data,
including independent ones
\cite{Mar02,Lev09,Ols07,Tot09b,Tur06,Tot07}. Currently the so
called thermal model is investigated but, up to now, it has not
given a solution to the riddle \cite{Tur09,Tur10}. For a relation
of the recent attempts to explain the anomaly, see \cite{Tur06},
Section 2.3, \cite{Tur06a}, Section 2 or \cite{Tur10b}, Section 6.
Up to now and more than thirty years after its discovery, the
Pioneer anomaly remains without a generally accepted solution,
even though it happens in our backyard, the solar system.

Note, however, that this work's arguments, based on the principle
of parametric invariance, give a solution of the riddle. In fact,
the authors of this paper proposed a model to explain this
intriguing phenomenon \cite{RT08,Ran04}, in which the
non-equivalence of $t_{\rm atom}$ and $t_{\rm astr}$ is due to the
combination of the fourth Heisenberg  relation and the unavoidable
coupling between the quantum vacuum and the background
gravitational potential $\Psi_{\rm bg}(t)$ that must pervade the
universe. The acceleration of the clocks is given in that model as
$a=\eta\dot{\Psi}_{\rm bg}(t)$, where $\eta$ is  a pure number
related to the electromagnetic properties of empty space and the
overdot means time derivative. However, as the presence of
$\Psi_{\rm bg}(t)$ indicates, that previous model is objectionable
since the very idea of potential is not well defined in general
relativity and cosmology, except in some cases as an
approximation. On the other hand, the arguments used in this work
do not use the concept of potential. They are based instead on the
principle of parametric invariance which is a very fundamental
principle in classical dynamics.

Nevertheless, the previous work \cite{RT08,Ran04} has some
interesting features. It can be applied to a limited region of
space, using the potential of the nearby bodies, not the
background one. Moreover it suggests a mechanism to explain the
physical reasons of the time acceleration. In fact, it is clear
that gravity surely affects the value of $t_{\rm astr}$, while the
quantum vacuum does not, the opposite being true for $t_{\rm
atom}$.
 One example of the application of the previous
model is given in reference \cite{RT09}, where it is shown that it
is fully compatible with the cartography of the solar system.

\section{Conclusions}
1. Building on the principle of parametric invariance, it was
shown that the concept of time is much more complex than is
usually assumed. It is important, in particular, to distinguish
between parametric time and dynamical time and to understand that
two stable, accurate and good but different clocks can be
non-equivalent. By this we mean that the times they measure could
accelerate with respect to one another if they are based on
different physical theories, as happens in the case of atomic  and
astronomical times, $t_{\rm atom}$ and $t_{\rm astr}$, which are
based on classical gravity and quantum electromagnetism,
respectively. This could be stated by saying that the principle of
parametric invariance has room for non-equivalent clock-times.

2. It is very important to understand that the description of a
dynamical system can not be considered complete without the
explicit mention of the chosen physical clock. This is specially
true in cosmology problems.

3. Although these arguments might seem rather formal, they are
also of practical importance. In particular, this work proposes an
explanation of  the Pioneer anomaly that is a refinement of a
previous one  and is fully compatible with the cartography of the
solar system \cite{RT08,RT09}. It is based on the non-equivalence
of the atomic time and the astronomical time, which happens to
have the same observational fingerprint as the anomaly. The
inverse time $a$ that characterizes the observations turns out to
be the second derivative of $t_{\rm atom}$ with respect to $t_{\rm
astr}$.

{\bf Acknowledgements} We are indebted to Profs. A. I. G. de
Castro, J. Mart\'in and J. Us\'on for discussions.

\end{document}